\documentclass[keywords]{revtex4}
\usepackage{graphicx}
\usepackage{dcolumn}
\usepackage{amsmath}
\usepackage{color}
\usepackage{epstopdf}
\usepackage{float}
\usepackage{graphicx}
\usepackage{subfigure}
\usepackage[utf8x]{inputenc}
\definecolor{coolblack}{rgb}{0.0, 0.18, 0.39}
\definecolor{darkred}{rgb}{0.5,0,0}
\definecolor{darkgreen}{rgb}{0,0.5,0}
\definecolor{darkblue}{rgb}{0,0,0.5}
\definecolor{lapislazuli}{rgb}{0.15, 0.38, 0.61}
\definecolor{venetianred}{rgb}{0.78, 0.03, 0.08}
\definecolor{bleudefrance}{rgb}{0.19, 0.55, 0.91}
\definecolor{dogwoodrose}{rgb}{0.84, 0.09, 0.41}
\usepackage[linktocpage,colorlinks]{hyperref}
\hypersetup{colorlinks=true, citecolor=darkgreen, linkcolor=blue,urlcolor = darkblue}
\makeatletter
\def\btt#1{\texttt{\@backslashchar#1}}
\DeclareRobustCommand\bblash{\btt{\@backslashchar}} \makeatother

\begin{document}
\title{Deflection of Light Around a Rotating BTZ Black Hole}
\author{Shubham Kala $^{a}$}\email{shubhamkala871@gmail.com}
\author{Hemwati Nandan$^{a,b}$}\email{hnandan@associates.iucaa.in}
\author{Prateek Sharma $^{a}$}\email{prteeksh@gmail.com}
\affiliation{$^{a}$Department of Physics, Gurukula Kangri Vishwavidyalaya, Haridwar 249 404, Uttarakhand, India}
\affiliation{$^{b}$Center for Space Research, North-West University, Mahikeng 2745, South Africa}

\begin{abstract}
\noindent We present a detailed study of gravitational lensing around a  rotating BTZ black hole in (2+1) dimensional gravity. The study of orbits for massless test particle around this BH spacetime is performed to  describe the nature of cosmological constant in lower dimensions. We study the effect of cosmological constant on the photon orbit in view of other critical parameters. The bending angle of light is studied in view of different values of cosmological constant for direct and retrograde motion of test particles.  It is being observed that the bending angle slightly decreases as the value of cosmological constant increases in the negative region. \\

\noindent keywords: Black Hole; Cosmological Constant; Gravitational Lensing.

\end{abstract}

\maketitle

\section{\normalsize Introduction}
{\normalsize
	\noindent 
	
	In general relativity (GR) a BH solution in $(2+1)$ dimensions with a negative cosmological constant is discovered by Bañados-Teitelboim-Zanelli (BTZ) in 1992 popularly known as BTZ BH spacetime \cite{banados1992black,Banados:1992gq}. It has opened a new window of opportunity for the physicists to study the nature of gravity in lower dimensions. This BH solution is different from conventional $ (3+1) $ dimensional BH solutions such as Schwarzschild and Kerr in many aspects, such as it is asymptotically anti de sitter instead of asymptotically flat and it has no curvature singularity.  Despite the absence of a curvature singularity at the origin, it has an event horizon, a Hawking temperature and other interesting thermodynamic properties \cite{quevedo2009geometric}. Moreover, in (2+1) dimensional GR, the Newtonian limit doesn't exist and there is no propagating degree of freedom.  However, this solution is regarded as a genuine solution due to the presence of event horizon, Hawking radiation and it also played an important role in the development of string theory \cite{Skenderis:1999bs,Peet:1997es,Polchinski:1995mt}. Furthermore, despite of its simplicity, the BTZ BH also plays a significant role to understand many physical properties in higher dimensions by using many toy models\cite{Bengtsson:2016dac,cvetivc2012graphene}. BTZ BH solution can be parameterized by two parameters mass and angular momentum \cite{Carlip:1995qv} and to understand the physical properties of a BH, study of null geodesics plays a significant role \cite{cruz1994geodesic,dasgupta2012geodesic}. Also, to understand the kinematics of photon around a (2+1) dimensional BH, the gravitational lensing (GL) can be regarded as an important technique. In order to study the optical observations of BHs through background light emission, photon sphere has been widely studied in its various aspects \cite{iyer2007light}. Stability of photon sphere plays a crucial role to explain the astrophysical phenomena such as deflection of light and BH shadows.   \\ 
	
	Over the last few decades, the GL is studied by many physicists due to its significance in astrophysics. Soldner was the first person who calculated the bending angle of light by using Newtonian Mechanics in 1801 \cite{jaki1978johann}. In 1959, Darwin calculated the light deflection angle due to a strong gravitational field using the Schwarzschild metric\cite{darwin1959gravity}. Another significant work involved the deflection angle and intensities for the images formed due to the Schwarzschild BH in terms of elliptic integrals of the first kind \cite{Uniyal:2018ngj,Sharma:2019qxd} . Considering the Schwarzschild BH for the strong GL, Virbhadra and Ellis obtained the lens equation and introduced a method to calculate the bending angle\cite{Virbhadra:1999nm}. Bozza treated the strong lensing phenomenon by a spherically symmetric BH, where an infinite sequence of higher order images are formed \cite{Bozza:2009yw} and later on extended for a spinning BH \cite{Bozza:2002zj}.\\
	
	In particular, the interpretation of the cosmological constant $(\Lambda)$ is a very fascinating and traditional topic in theoretical physics. On the observational side, large scale structure observations have made strong evidences for $\Lambda$ as a possible choice for dark energy. Since the cosmological constant should take part in all kinds of gravitational phenomena, investigations have been performed on different length-scales. From this point of view, the possible contribution of $\Lambda$  to the local bending of light have been extensively studied \cite{arakida2012effect,ishak2008new}. Even though such an effect would be many orders of magnitude too small to be measured with presently available instruments, it might in principle constitute one of the distinguishing characteristics of  $\Lambda$. The various authors have studied and obtained different kind of result on the influence of cosmological constant on light bending. The argument for the non-influence of $\Lambda$  was apparently first made in \cite{islam1983cosmological} and has been remade and reaffirmed by other authors, see for example \cite{rindler2007contribution,kerr2003standard,kagramanova2006solar,Arakida:2011ty}. Rindler and Ishak remove this contrary and provides a long over-
	due correction to previous works on whether or not the cosmological constant  affects the bending of light by a concentrated spherically symmetric mass \cite{Rindler:2007zz,Ishak:2008zc}. Here, we study the effect of cosmological constant in a non-spherical symmetric solution which will indeed provide a new insight the influence of $\Lambda$ on trajectories of photons. Based on above stand points we also study the GL around the fascinating BTZ BH and derive an exact expression for bending of light \cite{Iyer:2009wa,Bozza:2009yw,Virbhadra:1999nm} for more detailed view of the role of the cosmological constant in the deflection of light around a BH in lower dimensional gravity.\\
	
	This paper is organized as follows: In section 2, we introduce the metric and using Euler-Lagrange equation the geodesic equations are obtained. The study of variation of effective potential with BH parameters is presented accordingly. In section 3, we discuss critical variables and photon orbits along with present the existence of unstable photon orbits. In section 4, we have obtained an exact expression of bending of light for BTZ BH and analyzed its role in view of concerned parameters. We conclude our results in section 5.
	
	\section{\normalsize The BTZ BH Spacetime}
	The spacetime for (2+1) dimensional BTZ BH  is as follows \cite{Banados:1992gq},
	\begin{equation}
		ds^2 = -(N^2- r^2 N_\phi^2)dt^2 + \frac{1}{N^2} dr^2 + r^2d\phi^2 + 2r^2N_\phi dt d\phi.\label{metric}
	\end{equation}
	The square lapse $N^2$ and the angular shift  $ N_\phi$ in equation \ref{metric} are given by,\\
	$N^2 = (-M + \frac{r^2}{l^2} + \frac{J}{4r^2})$,   \hspace{0.5cm} $ N_\phi = \frac{-J}{2r^2}$.\\
	
	\noindent where $M$ and $J$ is the mass and  angular momentum respectively, while the radius of curvature is relates to cosmological constant as $l$ define as , $l=-(\Lambda)^\frac{-1}{2}$, which provides the necessary length scale in order to have a horizon of BH in 2+1 dimensions. The lapse function vanishes for two values of $r$ given by, 
	\begin{equation}
		r_{\pm} = l M^{1/2} \left[\frac{1}{2}\left(1 \pm \sqrt{1-\left(\frac{J}{M l}\right)^2}\right)\right]^{1/2},
	\end{equation}
	where $r_{+}$ is the BH horizon. From the Euler-Lagrange equation the generalized momentum for null geodesics can be written as,
	\begin{equation}
		\dot{t} = \dfrac{E r^{2} - L J /2}{r^{4}/l^{2}-M r^{2} + J^{2}/4},
	\end{equation}
	\begin{equation}
		\dot{\phi} = \dfrac{E J/2 - L M + L r^{2}/l^{2}}{r^{4}/l^{2}-M r^{2} + J^{2}/4}, 
	\end{equation}
	\begin{equation}
		\dot{r^{2}} = \dfrac{L^{2} M - E L J}{r^{2}} + E^{2} - \dfrac{L^{2}}{l^{2}}. \label{rdot2}
	\end{equation}
	Here, $E$ is the total energy of test particles at infinity while $L$ is the angular momentum of test particle. The radial equation is a quadratic function of $E$, and can be written as,
	\begin{equation}
		\dot{r}^2 = \left(E - V^{+}_{eff}\right)\left(E - V^{-}_{eff}\right),
	\end{equation}
	where the roots of $V^{+}_{eff}$ are given as follow,
	\begin{equation}
		V^{\pm}_{eff} = \frac{Jl}{2r^2} \pm \frac{1}{2} \sqrt{\frac{J^2 L^2}{r^4}-4\left(\frac{M L^2}{r^2}-\frac{L^2}{l^2}\right)}. \label{veff}
	\end{equation}
	Both the roots coincide at the event horizon. The expression of effective potential for BTZ BH is useful to study the photon orbits i.e., motion of photons, which had been studied in detail by Cruz et.al.\cite{cruz1994geodesic}. 
	\begin{figure}[H]
		\begin{center}
			\subfigure[]{\includegraphics[width=6.29cm,height=5cm]{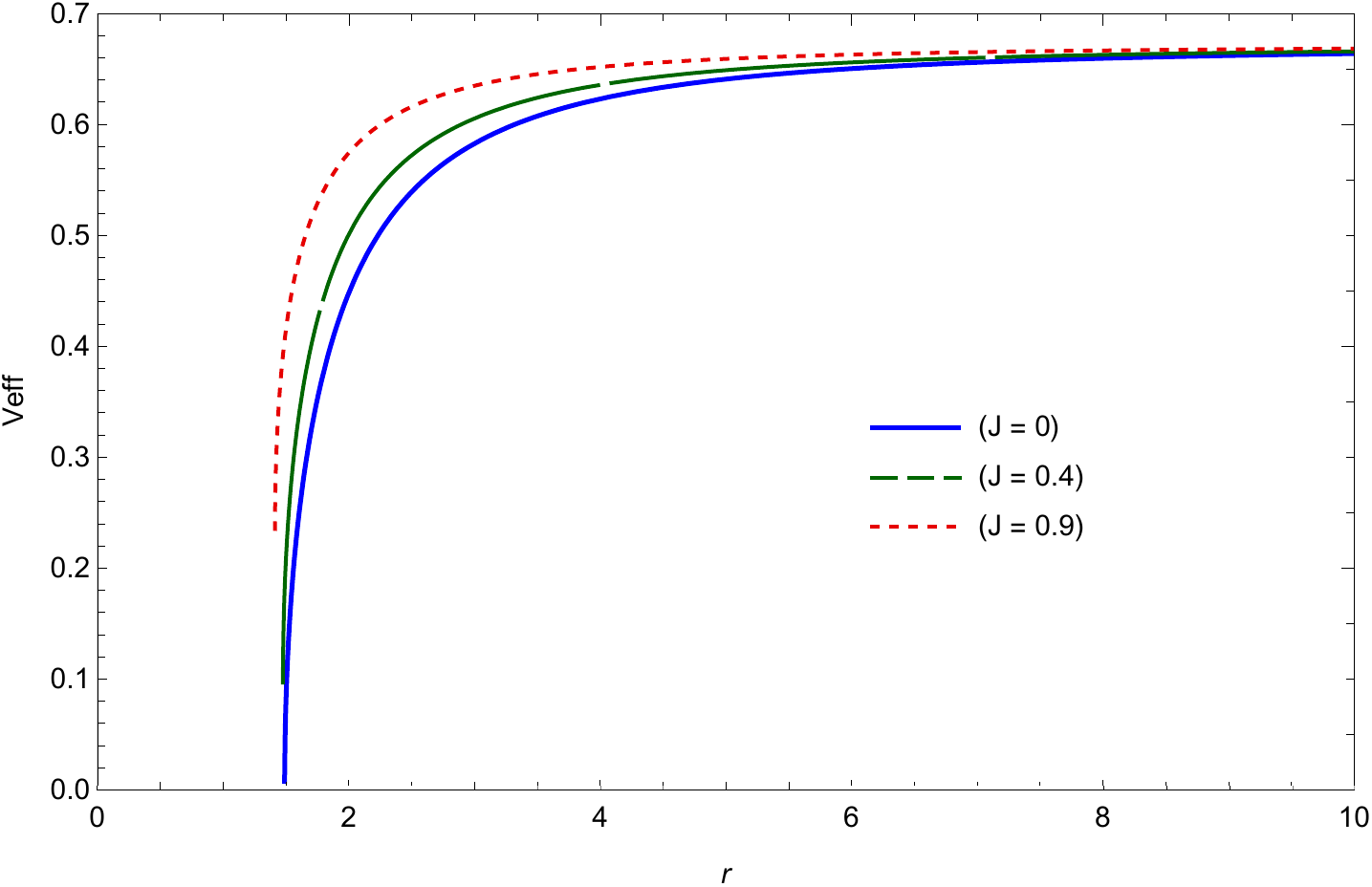}}
			\subfigure[]{\includegraphics[width=6.29cm,height=5cm]{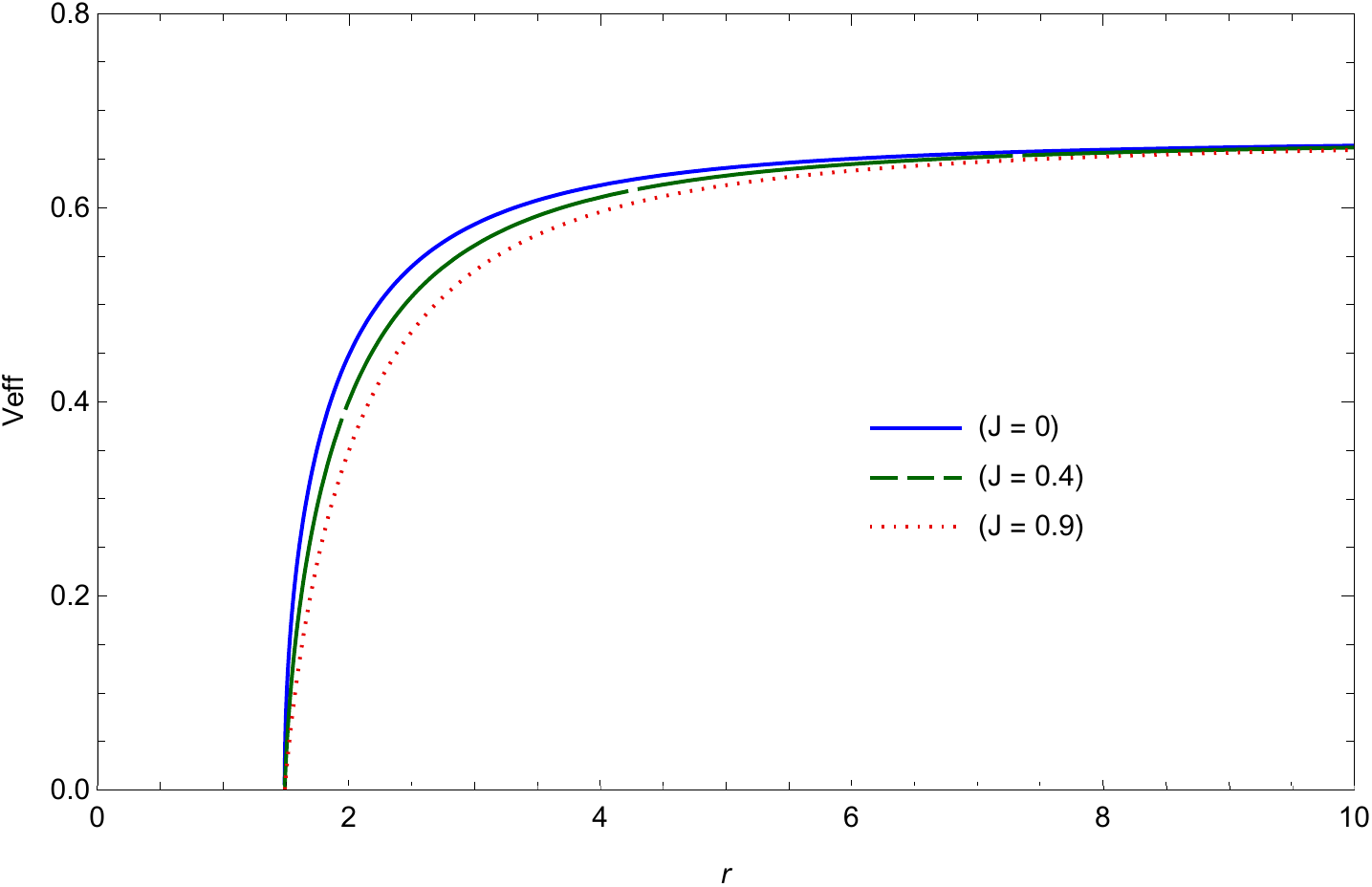}}
			\caption{Variation of effective potential with distance $r$ for different values of $ J $ when $M=1$ (a) $JL > 0$ (b) and $JL<0$}
		\end{center}
	\end{figure}
	\begin{figure}[H]
		\begin{center}
			\subfigure[]{\includegraphics[width=6.29cm,height=5cm]{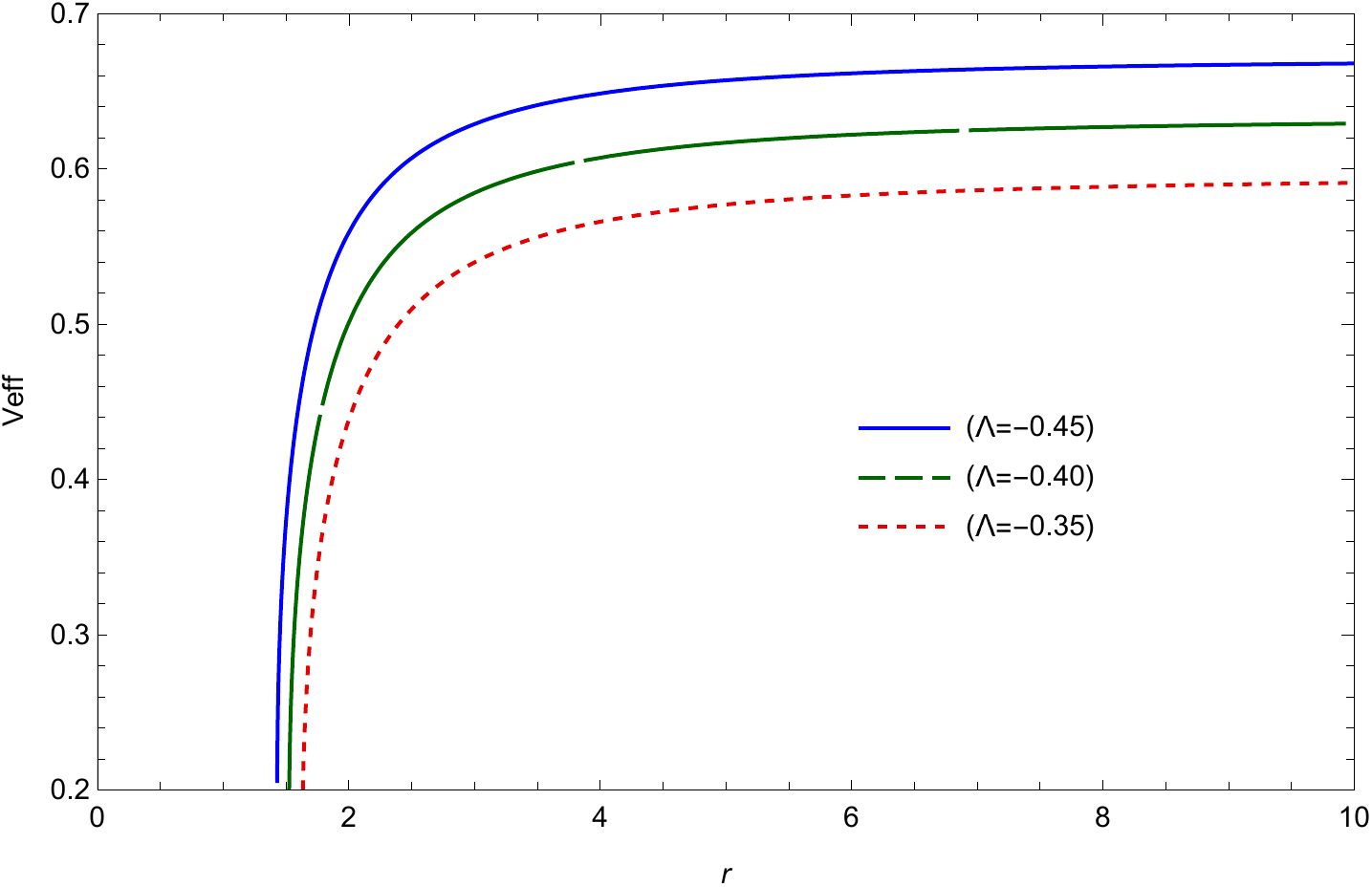}}
			\subfigure[]{\includegraphics[width=6.29cm,height=5cm]{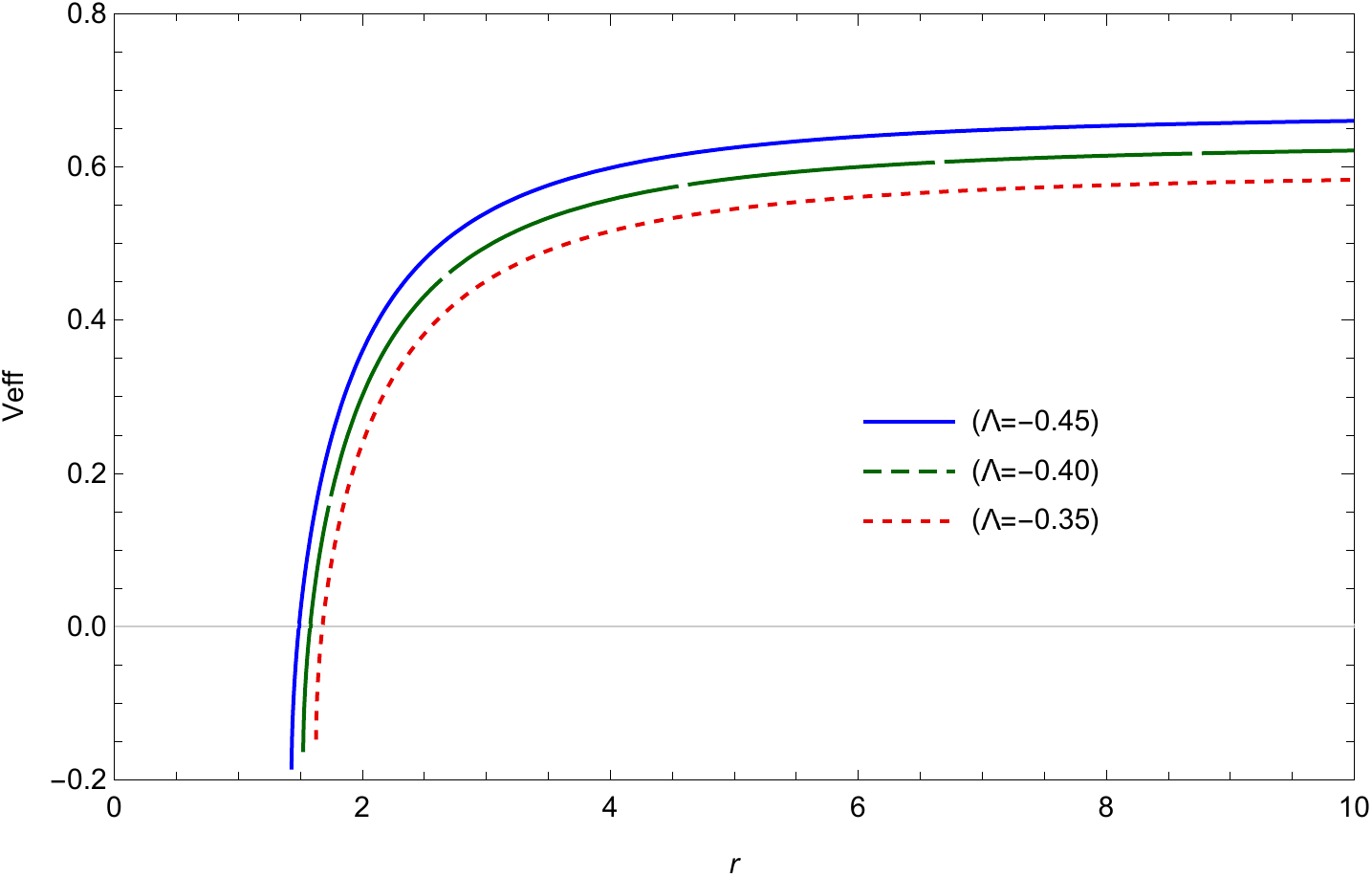}}
			\caption{Variation of effective potential with distance $r$  for different values of $ \Lambda $ when $M=1$ (a) $JL > 0$ (b) and $JL<0$}.
		\end{center}
	\end{figure}
	\noindent In Fig(1) and Fig(2), the effective potential is presented graphically  to study the behavior of photon near BH for different values of $J$ and $l$ respectively. In each curve, there is no minima and therefore no stable orbit for the photons exists, only an unstable orbit exists in each case which corresponds to the maximum value of effective potential described by equation \ref{veff}. \\
	
	Substituting $ u = 1/r $, we obtain,
	\begin{equation}
		\dfrac{d\phi}{du} = \dfrac{\left(E \frac{J}{2} - L M\right) u ^{2} + \frac{L}{l^{2}}}{\left(\frac{J^{2} u^{4}}{4} - M u^{2} + \frac{1}{l^{2}}\right)} \cdot \dfrac{1}{\sqrt{\left(E^{2} - \frac{L^{2}}{l^{2}}\right) - \left(E L J - L^{2} M\right) u ^{2}}}, \label{eom}
	\end{equation}
	The above expression \ref{eom} can be simplified as,
	\begin{equation}
		\dfrac{d\phi}{du} = \dfrac{C_{+}}{\eta\hspace{0.1cm} u_{+} \left(1 - m u^{2}\right)\sqrt{1-k u^{2}}} + \dfrac{C_{-}}{\eta \hspace{0.1cm} u_{-} \left(1 - n u^{2}\right)\sqrt{1-k u^{2}}}. \label{eom2}
	\end{equation}
	The variables in the equation \ref{eom2} are defined as follows,\\
	\begin{center}
		$ C_{\pm} = \mp \left[\dfrac{L J^{2}  + 2 l^{2}\left(\frac{E J}{2} - L M\right)\left(M \pm \gamma \right)}{J^{2} l^{2} \gamma}\right]$,\\ $ u_{\pm} = \frac{2}{J^{2}}, \left(M\pm\gamma\right) $, $ \gamma = \sqrt{M^{2} - \frac{J^{2}}{l^{2}}} $,
		$ \eta = \sqrt{E^{2} - \frac{L^{2}}{l^{2}}} $,\\ $ m = 1/ u_{+} $,  $ n = 1/u_{-} $  and $ k = \dfrac{E L J - L^{2} M}{E^{2} - \frac{L^{2}}{l^{2}}} $.\\
	\end{center}
	The distance of closest approach may then easily be obtained by equation \ref{rdot2} and can be written as,
	\begin{equation}
		r_{0} = \sqrt{\frac{ELJ-ML^2}{E^2-\frac{L^2}{l^2}}}. 
	\end{equation}
	\begin{figure}[H]
		\begin{center}
			\includegraphics[width=10cm,height=7cm]{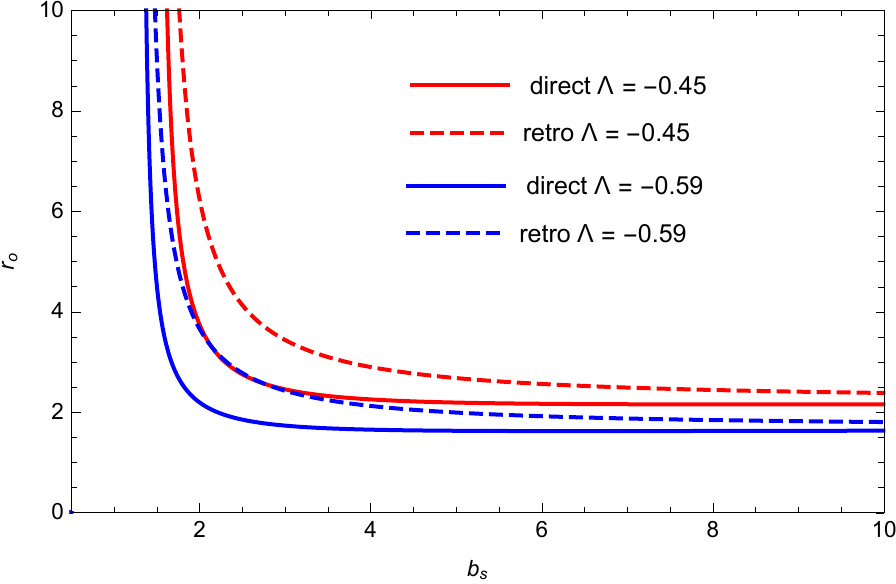}
			\caption{Variation of distance of closest approach with impact parameter.}
		\end{center}
		
	\end{figure}
	
	\section{\normalsize Critical Variable and Photon Orbit}
	The massless particle moves in a closed orbit around a BH due to the curvature of BH spacetime. Such a geodesic is generally known a photon orbit and a photon orbits are in general unstable a slight perturbation can either cause the particle to plunge into the BH, or cause it to escape to the infinity \cite{Tang:2017enb,Khoo:2016xqv}. Since the BTZ BH solutions are asymptotically AdS with negative cosmological constant and it is therefore, important to study the behavior of the photon orbits in such lower dimensional BH spacetime. The existence of a photon sphere in this particular spacetime has also important implications for GL. In any BH spacetime containing a photon sphere, various aspects of GL will give rise to relativistic images and in order to investigate aspects of GL, the metric coefficient $N^2=f(r)$ leads to,
	\begin{equation}
		f'(r) = \frac{2r}{l^2}-\frac{J^2}{2r^3}
	\end{equation}
	We now consider the extremal case and setting both the metric coefficient and its derivative to be zero, such that
	\begin{equation}
		J_{c} = m_{c}l, r_{c} = \sqrt{\frac{J_{c} l}{2}} = l \sqrt{\frac{m_{c}}{2}}.
	\end{equation}
	The effective potential can also be written as in terms of $b(=\frac{L}{E})$, where $L$ and $E$ denote the angular momentum and the energy of the massless particle respectively as below,
	\begin{equation}
		V_{eff} = \frac{N^2}{r^2} - \frac{2 N^{\phi}}{b} - N{^2{\phi}} = \frac{1}{r^2}\left(\frac{r^2}{l^2} - m + \frac{J}{b}\right)
	\end{equation}
	In the extremal case, the effective potential and its derivative then read as,
	\begin{equation}
		V_{eff} = \frac{-1}{r^{2}}\left(m_{c} - \frac{m_{c} l}{b} - \frac{r^{2}}{l^{2}}\right), \hspace{2cm} V'_{eff} = \frac{-2 m_{c}}{r^3}\left(\frac{l}{b}-1\right) \label{V_{eff}}.
	\end{equation}
	Thus, $b=l$ is the only one solution for which $V_{eff}=0$. The surface plot in \figurename{\ref{sp}} indicate that in this case, the effective potential is always a constant i.e. $V_{eff} = b^{-2} = l^{-2}$. A constant potential means that a particle can easily move across it, and therefore is extremely unlikely to remain in any fixed orbit for long. Beyond the horizon, the effective potential becomes constant and there is no intersection with vertical plane, therefore, there is only unstable photon orbits which totally depends on $b$. Thus for extremal BTZ BH, one can have photon orbit when $b=l$ and these photon orbits are all 'unstable'. One may also notice that from \figurename{\ref{sp1}} and \figurename{\ref{sp2}} there are no changes on the behavior of effective potential for the different values of cosmological constant i.e. $\Lambda=-0.45$ and $\Lambda=-0.40$ respectively.
	
	\begin{figure}[H]
		\begin{center}
			\subfigure[]{\includegraphics[width=7cm,height=5cm]{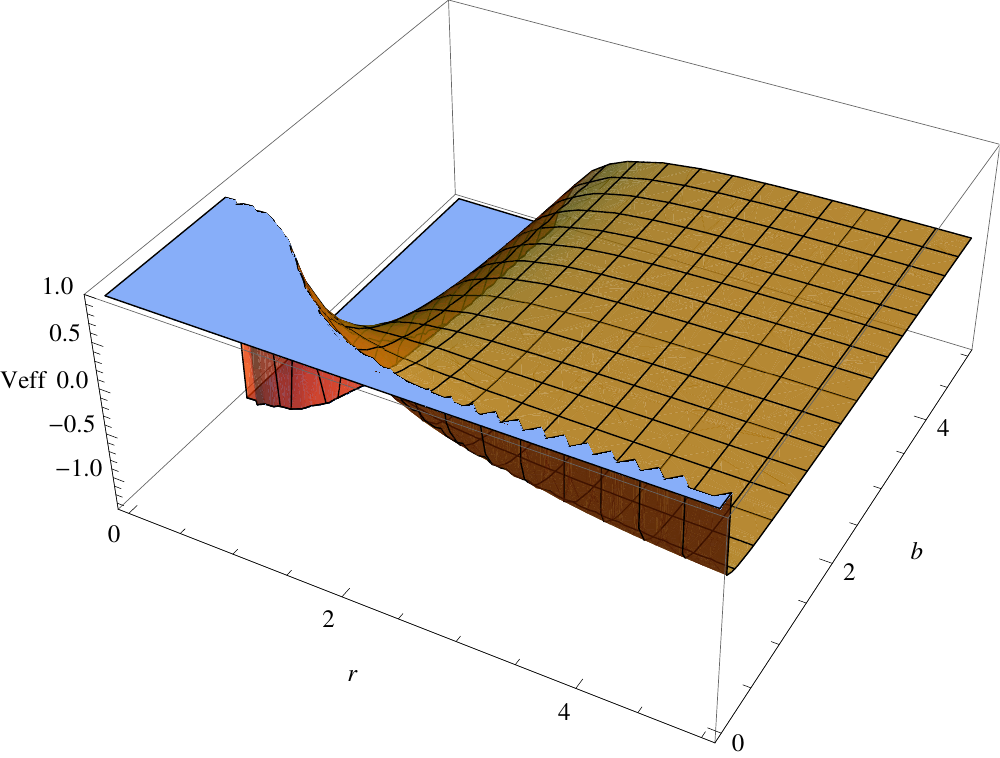}\label{sp1}}
			\subfigure[]{\includegraphics[width=7cm,height=5cm]{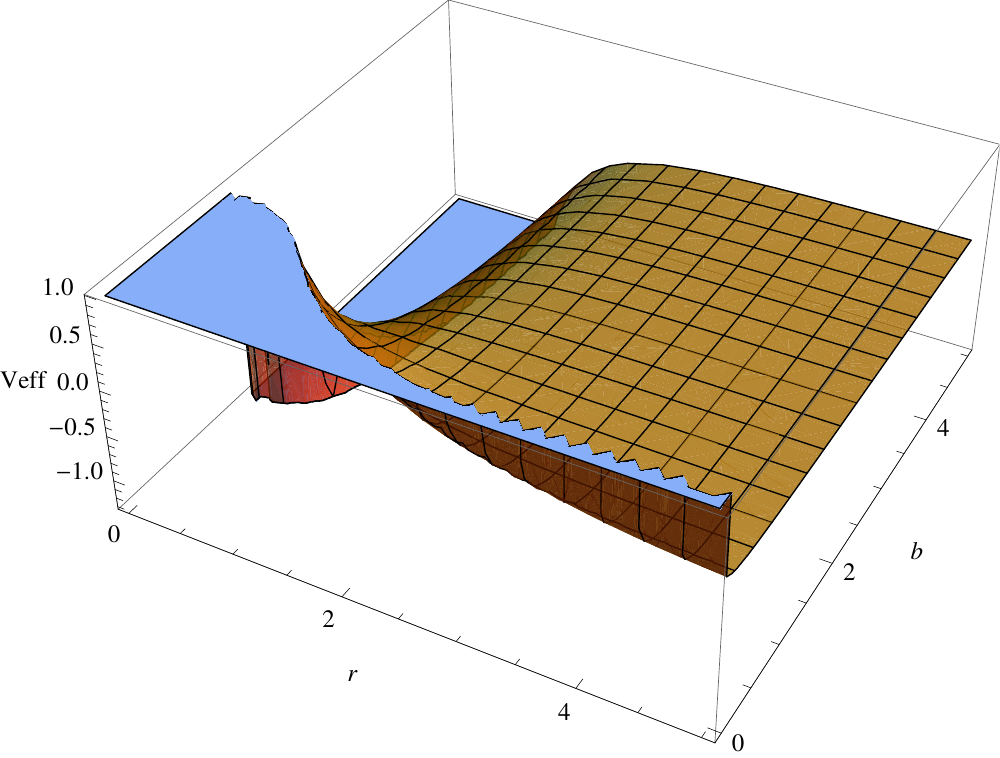}\label{sp2}}
			\caption{The surface plot of the first derivative of the effective potential $V_{eff}$, of an extremal BTZ BH, with mass $m_{c}$ set to unity, as a function of the impact parameter $b$, and radial distance $r$. Here, the radius of curvature (a) $l=1.49(\Lambda=-0.45)$ and (b) $l=1.58(\Lambda=-0.40)$.}  \label{sp}
		\end{center}
	\end{figure}\label{sp}
	
	\section{\normalsize The Exact  Expression For Bending Angle Of Light}
	Let us consider a light ray approaches, the BH such that it starts from the asymptotic region, with $r_{0}$ as the distance of closest approach and an emerging light ray reaches upto an observer in the asymptotic region. Now from the definition of deflection of light, one knows that the change in $\Phi$ and the bending angle $\delta$ simply related by a difference of $\pi$. Hence the expression of bending angle for BTZ BH can be defined as \cite{PhysRevD.80.124023,Uniyal:2018ngj,Sharma:2019qxd},
	\begin{equation}
		\delta = 2\int_{0}^{u_{0}} \left[\dfrac{C_{+}}{\eta\hspace{0.1cm} u _{+} \left(1- m u^{2}\right) \sqrt{1 - k u^{2}}} + \dfrac{C_{-}}{\eta\hspace{0.1cm} u _{-} \left(1- n u^{2}\right) \sqrt{1 - k u^{2}}}\right] du - \pi
	\end{equation}
	We have solved the above indefinite integral which leads to, 
	\begin{multline}
		\delta = \dfrac{2C_{+}}{\eta\hspace{0.1cm} u_{+} \sqrt{m-k}}\tanh^{-1}\left[\dfrac{u \sqrt{m-k}}{\sqrt{1-k u^{2}}}\right]\Biggr|_{0}^{u_{0}}  \\ +  \dfrac{2C_{-}}{\eta\hspace{0.1cm} u_{-} \sqrt{n - k}}Tanh^{-1}\left[\dfrac{u \sqrt{n-k}}{\sqrt{1-k u^{2}}}\right]\Biggr|_{0}^{u_{0}}  - \pi
	\end{multline} 
	where,\\ $u_{0} = \sqrt{\frac{E^{2}- \frac{L^{2}}{l^{2}}}{E L J - M L^{2}}},$ which is inverse of the distance of closest approach. \\
	
	Using the hyperbolic inverse expansion function, the exact expression of bending angle reduce to,
	\begin{equation}
		\delta = \frac{C_{+}}{\eta\hspace{0.1cm} u_{+}\sqrt{m-k}}\ln (2 u_{0} \sqrt{m-k}) + \frac{C_{-}}{\eta\hspace{0.1cm} u_{-}\sqrt{n-k}}\ln (2 u_{0}\sqrt{n-k}) - \pi.
	\end{equation}\\
	
	The plots in \figurename{\ref{fig:bb}} represents the variation of bending angle with inverse of the distance for closest approach and the cosmological constant. The bending angle for direct motion increases as the distance of closest approach decreases and the same trend is observed for the retrograde motion. It is observed that the bending angle decreases as the value of cosmological constant increases in negative region for direct and retrograde motion of photon. In the figure it is clear that the qualitative nature of bending angle does not change due to the direction of motion of photons and cosmological constant. However, there exist a marginal effect on the degree of bending of light around the BH from these parameters. 
	
	\begin{figure}[H] 
		\begin{center}
			\includegraphics[width=9.5cm,height=6.0cm]{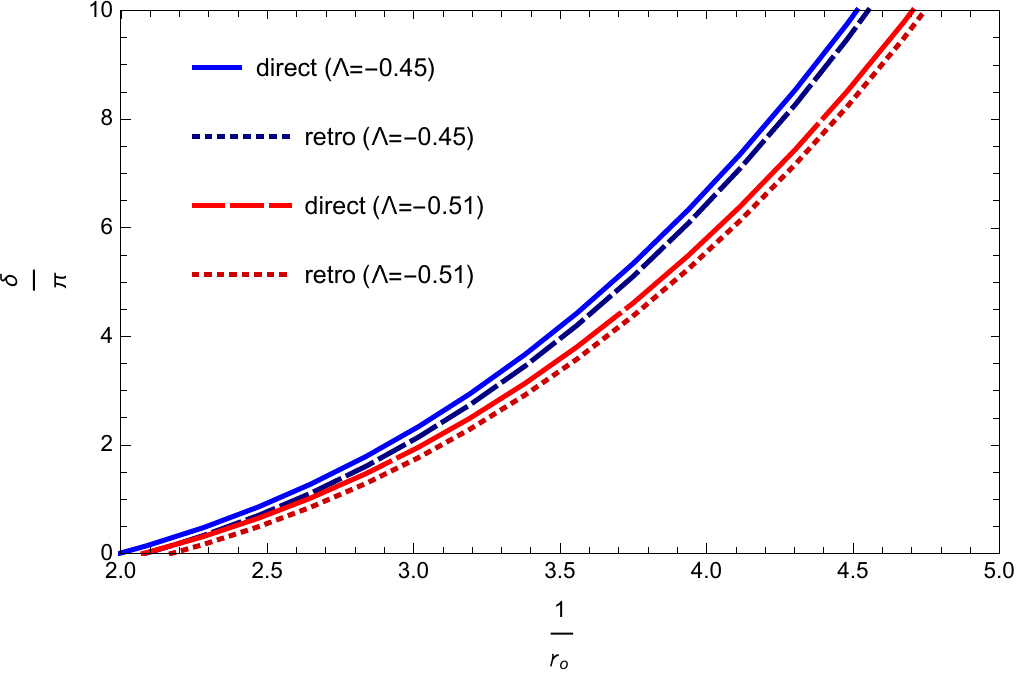}\label{bvsr0}
			\caption{The variation of bending angle with inverse of distance of closest approach when $ E = 2 $, $L=0.1$ for direct, $L=-0.1$  for retrograde motion and $ J=0.5 $. }\label{fig:bb}
		\end{center}
	\end{figure}

	\section{Conclusions}
	In this paper, we have studied the null geodesic equations for BTZ BH spacetime in (2+1) dimension through the investigation of variation of effective potential with the spin parameter $(J)$ and radius of curvature $(l)$. The value of effective potential for direct orbits $(i.e. JL>0)$ increases with spin parameter and radius of curvature. Also, the value of effective potential for retrograde orbits $(i.e. JL<0)$ decreases as the value of spin parameter with radius of curvature increases. The effective potential for direct motion of photons becomes constant after a critical value which suggest only unstable orbits are possible in this region. However, for retrograde motion the effective potential also lies in negative region therefore, the bound orbit may be exist in this region. The surface plots for effective potential indicates the behavior of potential is constant with radial distance and impact parameter. Also, the effective potential totally depends on impact parameter of BH spacetime and these orbits are unstable i.e. there are no stable photon orbits are possible near its horizon.\\
	One of the initial motivation for this study is to investigate a physical phenomenon like gravitational lensing in (2+1) dimensional BH spacetime. We have studied the bending angle of light with various parameters of BTZ BH and calculated exact bending angle of light along with its variation with distance of closest approach and cosmological constant. We observed that the variation of bending angle is conventional. It increases with decreases in distance of closest approach for direct and retrograde motion of photon for negative values of cosmological constant. However, it is noticed that as the value of cosmological constant increases in negative region the bending angle marginally decreases for a particular value of distance of closest approach.

	\section{Acknowledgments}
	The authors are thankful to Radouane Gannouji for discussions and useful suggestions during the initial stage of this work. The author SK acknowledge the financial support provided by the Uttarakhand State Council
	of Science and Technology (UCOST), Dehradun through the grant number $UCS\&T/R\&D-18/18-19/16038/4$. The authors HN, PS acknowledge the financial support provided by Science and Engineering Research Board (SERB), New Delhi through the Grant No. EMR/2017/000339. S. Kala, H. Nandan and P. Sharma also acknowledge the facilities at ICARD, Gurukula Kangri Vishwavidyalaya, Haridwar..\\
	
	\bibliographystyle{unsrt} 
	\bibliography{sample}
	
\end{document}